\documentclass[12pt,epsf]{article}
\usepackage{graphicx}

\usepackage{geometry}                
\usepackage{graphicx}
\usepackage{amssymb}
\usepackage{epstopdf}
\usepackage{slashed}
\usepackage{float}
\usepackage{hyperref}
\usepackage{amsmath}
\usepackage{chngcntr}
\usepackage{multirow}
\numberwithin{equation}{section}

\newcommand{\bea}{\begin{eqnarray}}
\newcommand{\eea}{\end{eqnarray}}
\newcommand{\beas}{\begin{eqnarray*}}
\newcommand{\eeas}{\end{eqnarray*}}
\newcommand{\ba}{\begin{array}}
\newcommand{\ea}{\end{array}}

\newcommand{\ppderivative}[2]{\frac{\partial^2 #1}{\partial #2^2}}
\newcommand{\psderivative}[3]{\frac{\partial^2 #1}{\partial #2\partial #3}}

\newcommand{\pare}[1]{\left( #1 \right)}
\newcommand{\colc}[1]{\left[ #1 \right]}
\def\be{\begin{equation}}
\def\ee{\end{equation}}
\def\ben{\begin{equation*}}
\def\een{\end{equation*}}
\def\beqa{\begin{eqnarray}}
\def\eeqa{\end{eqnarray}}

\renewcommand{\a}{\alpha}

\renewcommand{\b}{\beta}

\newcommand{\m}{\mu}
\newcommand{\n}{\nu}
\renewcommand{\r}{\rho}

\renewcommand{\d}{\partial}

\newcommand{\vect}[1]{\vec{#1}}
\def\be{\begin{equation}}
\def\ee{\end{equation}}

\def\beqa{\begin{eqnarray}}
\def\eeqa{\end{eqnarray}}

\def\bi{\begin{itemize}}
\def\ei{\end{itemize}}

\begin{document}

\begin{titlepage}
\hfill

\vspace*{20mm}
\begin{center}
{\Large \bf Density versus chemical potential in holographic probe theories}

\vspace*{15mm}
\vspace*{1mm}
Fernando Nogueira and Jared B. Stang\footnote{e-mails: nogueira@phas.ubc.ca, jstang@phas.ubc.ca}
\vspace*{1cm}

{\it Department of Physics and Astronomy,
University of British Columbia\\
6224 Agricultural Road,
Vancouver, B.C., V6T 1W9, Canada}

\vspace*{1cm}
\end{center}

\begin{abstract}
We study the relationship between charge density $(\rho)$ and chemical potential $(\mu)$ for an array of Lorentz invariant $3+1$ dimensional holographic field theories with the minimal structure of a conserved charge. In all cases, at large density, the relationship is well modelled by a power law behaviour of the form $\rho\propto\mu^\alpha$. For the minimal ingredients of a gravitational field and a probe $U(1)$ gauge field in the bulk, we find general constraints $\a=1$ for the Maxwell action and $\a>1$ for the Born-Infeld case, for general background metrics. We show that the constraint $\a\ge1$ can also be understood directly in the field theory from thermodynamic stability and causality. We then determine which values of $\a$ are realized in a large range of example systems, including D$p$-D$q$ probe brane constructions and `bottom-up' models with gauge and scalar fields.
\end{abstract}
\vskip 2cm

\begin{center}
\vskip 1cm

\end{center}
\end{titlepage}

\newpage
\tableofcontents

\section{Introduction}

The AdS/CFT correspondence \cite{Maldacena:1997re, Gubser:1998bc, Witten:1998qj}, which conjectures the equivalence of a gravity theory in $d+1$ dimensions and a gauge theory in $d$ dimensions, has become a valuable tool for the study of strongly coupled field theories. Using the correspondence, many questions about quantum field theories may be phrased in the context of a gravity theory; in the limit of strong coupling, certain previously intractable field theory calculations are mapped to relatively simple classical gravity computations. 

\subsubsection*{Holography and finite density}

One difficult regime of strongly coupled field theory that gauge / gravity duality is particularly suited to study is that of finite charge density. Here, lattice techniques fail due to the `sign problem': at finite chemical potential, the Euclidean action becomes complex which results in a highly oscillatory path integral. We can avoid this difficulty by mapping the problem to a gravity dual using the AdS/CFT dictionary. According to the dictionary, in order to have a global $U(1)$ symmetry in the field theory, one needs to include a $U(1)$ gauge field in the gravity bulk. The charge density and chemical potential are encoded in the asymptotic behaviour of the gauge field. At strong coupling in the field theory, the bulk theory is well described by classical gravity, and one may solve the classical equations of motion on the gravity side to study the field theory at finite density.

Given this relatively simple access to finite density configurations, we might hope that some physically realistic strongly interacting systems may be approximately described by a holographic dual. In this case, qualitative features of the holographic theory would carry over to the exact theory. It would be useful to characterize the types of finite density field theories that have a dual formulation and admit this type of study.

In this paper, we seek to answer this question from the perspective of the holographic theory. Specializing to holographic probes, in which fields are considered as small fluctuations on fixed gravitational backgrounds, we study systems with the minimal structure of a conserved charge and find the $\rho-\mu$ relations that are possible in the field theory duals. We attack this problem by first deriving constraints on the relationship based on general grounds before studying several specific examples of holographic field theories.

\subsubsection*{Summary of results}

In our study, we observe that, at large densities, the field theory dual to a substantial class of gravity models can be described by a power law relation of the form\footnote{Here and throughout, $\alpha$ refers to the power in this form of $\rho-\mu$ relationship.}
\begin{equation}\label{powerlaw}
\rho=c\mu^\alpha.
\end{equation}

Firstly, we look to understand the constraints on the the $\rho-\mu$ relationship from the point of view of the field theory, using local stability and causality. Usually, results here depend on the particular form of the free energy. In all cases with $\rho-\mu$ behaviour (\ref{powerlaw}), local thermodynamic stability places the condition $\alpha>0$ on the exponent. In general, for a theory at low temperature, we may write the particular free energy expansion $f\propto- \m^{\a+1}-a\m^{\beta}T^\gamma$, with $\gamma>0$ and $a>0$, with corresponding charge density $\rho\propto (\a+1)\m^\a+a\b\m^{\b-1}T^{\gamma}$. Combined, local stability and causality demand that $\a\ge 1$ and $\gamma>1$.

Next, we consider Born-Infeld and Maxwell actions for the gauge field in a generic background. Under mild assumptions, in both cases, the power $\alpha$ is constrained. For the Born-Infeld action, the condition \be\a>1\;\;\;\;\;\;\;\; \textrm{(Born-Infeld action)}\ee arises,\footnote{Naively, we could construct systems for which $\alpha\leq1$, however, in these situations, the contribution of the constant charge density to the total energy diverges, consequently we can not say that there is a power law relation.  This divergence signals a breakdown of the probe approximation rendering these systems outside the scope of these notes. Notice that $\alpha>1$ is consistent with the bound derived from stability and causality.} while, for the Maxwell action, the power law coefficient is fixed to \be\a=1.\;\;\;\;\;\;\;\; \textrm{(Maxwell action)}\ee  Interestingly, these conditions are in agreement with those derived from field theory considerations, giving rise to the same range of possible values of $\a$. In summary, all power law relationships consistent with stability and causality can be realized in simple probe gauge field setups by varying the background metric.

To see which values of $\a$ arise for backgrounds corresponding to specific models, we explore a variety of $3+1$ Poincar\'{e}-invariant holographic field theories dual to D$p$-D$q$ brane systems and `bottom-up' models with gauge and scalar fields. The former have been used, for example, in studies of holographic systems with fundamental matter \cite{Sakai:2004cn,Mateos:2006nu,Nakamura:2006xk,Kobayashi:2006sb,VanRaamsdonk:2009gh}, producing many features of QCD, including confinement,\footnote{It was recently pointed out that the usual identification of the black D4 brane as the strong coupling continuation of the deconfined phase in the field theory is not valid \cite{Mandal:2011ws}.} chiral symmetry breaking, and thermal phase transitions \cite{Kruczenski:2003uq,Babington:2003vm,Aharony:2006da,Bergman:2007wp}. Bottom-up, phenomenological models have been studied in various model-building applications including superconductors\footnote{A top-down realization of a gauge / gravity superconductor has been found in \cite{Ammon:2009fe}.} \cite{Hartnoll:2008kx,Hartnoll:2009sz,Horowitz:2010gk,Hartnoll:2008vx,Horowitz:2008bn,Horowitz:2009ij} and superfluids \cite{Herzog:2008he,Basu:2008st,Arean:2010zw}.

In the D$p$-D$q$ systems, table \ref{summary1}, a variety of powers $\alpha$ in the range $1<\a\le3$ are realized, respecting the $\alpha>1$ constraint. Note that these results only involve the Born-Infeld action and neglect couplings of the brane to other background spacetime fields.

\begin{table}[H]
\begin{center}
\begin{tabular}{ c | c  c  c  c  c  c | c c c |} 
\cline{2-10}
& \multicolumn{9}{|c|}{Probe brane}\\
\cline{2-10}
& \multicolumn{6}{|c|}{ $d=4$} & \multicolumn{3}{|c|}{$d=5$}\\
\hline 
\multicolumn{1}{|c|}{Background branes} & D9 & D8 & D7 & D6 & D5 & D4 & D8 & D7 & D6 \\
\hline
\multicolumn{1}{|c|}{D3} & 3 & & 3 & & 3 & & & &\\
\multicolumn{1}{|c|}{D4} & & 5/2 & & 2 & & $3/2$ & 3 & & 5/2 \\
\multicolumn{1}{|c|}{D5} & & & 2 & &  & & & 2 &\\
\multicolumn{1}{|c|}{D6} & & & & 3/2 & & & & &\\
\hline 
\end{tabular}
\caption{The power $\alpha$ in the relationship $\rho\propto\mu^\alpha$ at large $\rho$ for $3+1$ dimensional field theories dual to the given brane background with the indicated probe brane, with $d-1$ shared spacelike directions. For $d=5$ the theory is considered to have a small periodic spacelike direction while for background D$p$ branes with $p>3$, the background is compactified to $3+1$ dimensions.}\label{summary1}
\end{center}
\end{table}

In the phenomenological probe models, table \ref{summary2}, in all cases except one (the probe gauge field in the black hole background), the dominant power $\alpha$ is determined by conformal invariance, since we consider asymptotically AdS backgrounds.\footnote{Different power laws can arise for holographic theories on different backgrounds, such as Lifshitz spacetimes. However, these will not be considered here.} Since $\mu$ and $T$ are the only dimensionful parameters, the density must take the form $\rho=\mu^{d-1}h(T/\mu)$, where the underlying space has $d$ spacetime dimensions. At large $\mu$ and fixed $T$, we can expand $h$ to see that $\mu^{d-1}$ dominates the $\rho-\mu$ relationship. In systems with one small periodic spacelike direction, the dominant power $\alpha$ is larger than the corresponding theory without a periodic direction since, at large densities, on the scale of the distance between charges, the theory is effectively higher dimensional.\footnote{The phase transition that holographic theories with a periodic direction undergo as the density increases was studied in \cite{Kaplunovsky:2012gb}.} Our study of bottom-up models also includes an analysis of the gravity models in the full backreacted regime. As seen in table \ref{summary2}, the power law $\a$ in these cases is also determined by the same conformal invariance argument. 

In these bottom-up models we are more interested in the detailed behaviour at intermediate values of $\m$. It is found that, in general, when the scalar field condenses in the bulk, the corresponding field theory is in a denser state than that without the scalar field. As well, the field theory dual to the gauge field and scalar field in the soliton background is in a denser state than that dual to the same fields in the black hole background. In the systems with a scalar field, at large $\m$, the $\r-\m$ relationship is well fit by the form $\r=c(q,m^2)\m^\a$,\footnote{In the probe cases we can scale $q$ to 1, leaving $c=c(m^2)$.} where $q$ and $m^2$ are the charge and mass-squared of the scalar field. While the power $\a$ is fixed by the conformal invariance, we find that the scaling coefficient $c(q,m^2)$ increases with increasing $q$ or decreasing $m^2$.

\begin{table}[H]
\begin{center}
\begin{tabular}{| c | c | c || c | c |} \hline
Regime & Background & Fields & $d=4$ & $d=5$ \\ \hline
\multirow{3}{*}{probe} & \multirow{2}{*}{black hole} & $\phi$ & 1 & 1 \\
& & $\phi, \psi$ & 3 & 4 \\ \cline{2-5}
& soliton & $\phi, \psi$ &  & 4 \\ \hline
\multirow{3}{*}{backreacted} & \multirow{2}{*}{black hole} & $\phi$ & 3 & 4 \\ 
&  & $\phi, \psi$ & 3 & 4 \\ \cline{2-5}
& soliton & $\phi, \psi$ & & 4 \\ \hline
\end{tabular}
\caption{The power $\alpha$ in the relationship $\rho\propto\mu^\alpha$ at large $\rho$ for $3+1$ dimensional field theories dual to the given gravitational background with the stated fields considered in either the probe or backreacted limits. $\phi$ is the time component of the gauge field, $\psi$ is a charged scalar field, and $d$ is the number of spacetime dimensions. For $d=5$ the theory is considered to have a small periodic spacelike direction.}\label{summary2}
\end{center}
\end{table}

\subsubsection*{Organization}

In section \ref{bg}, we discuss some possible general examples of finite density field theories and attempt to establish bounds on the $\r-\m$ relationship by imposing thermodynamical constraints on these systems. In section \ref{gaugefield} we briefly introduce holographic chemical potential and find, for Maxwell and Born-Infeld types of action, under mild assumptions, to what extent they reproduce the relationship found in \ref{bg}. In section \ref{probes} we investigate the probe limit of both top-down and bottom-up theories; first we study D$p$-D$q$ systems, then we move to gauge and scalar fields in both black hole and soliton (with one extra periodic dimension) backgrounds. Section \ref{br} extends the analysis of the bottom-up models to include the backreaction of the fields on the metric.
\subsubsection*{Relation to previous work}

Some of the results presented in these notes have appeared previously in the literature. Finite density studies for probe brane systems have appeared for the Sakai-Sugimoto model \cite{Bergman:2007wp,Kim:2006gp,Horigome:2006xu,Rozali:2007rx}, the D3-D7 system \cite{Nakamura:2006xk,Kobayashi:2006sb,Erdmenger:2007ja,Erdmenger:2008yj,Ammon:2011hz}, and the D4-D6 system \cite{Matsuura:2007}. The bottom-up models we consider are naturally studied at finite chemical potential (see, for example, \cite{Hartnoll:2009sz} for the black hole case and \cite{Horowitz:2010jq} for the soliton dual to a $2+1$ dimensional field theory) due to the presence of the gauge field. 

Our work focusses on the $\rho-\mu$ relation at large chemical potential over a broad class of theories that are dual to $3+1$ dimensional field theories. We find, on very general grounds, constraints on the $\rho-\mu$ relation in holographic models constructed from Maxwell and Born-Infeld actions. Additionally, we use thermodynamical considerations to constrain the $\rho-\mu$ relation from the field theory point of view and find that these constraints are in agreement with those derived holographically. Further, we extend the analysis in the above references to the large density regime and include additional examples, collecting the results of a large range of models.


\section{CFT thermodynamics}\label{bg}

In this section, by appealing to local thermodynamic stability and causality in the field theory, we attempt to establish generic constraints satisfied by the coefficient $\a$ from a purely field theory stand point. 
The results found here will lay ground for our intuition when approaching this problem from the holographic side.

\subsubsection*{Generic system at large chemical potential}

In order to study the density and chemical potential from the field theory perspective, we begin with a general ansatz for the free energy of a hypothetical system. In the large density limit, we expect that the chemical potential will dominate the expression, so we may write\footnote{Recall $\rho=-(\d f/\d\mu)_T$ so that, again, $\rho\propto\mu^\alpha$.}
\be\label{ansatz}
f\propto- \m^{\a+1}-a\m^{\beta}T^\gamma+\dots,
\ee
where the dots denote corrections higher order in $T/\mu$. For $a$ positive, imposing a positive entropy density $s=-(\d f/\d T)|_{\m}>0$ implies $\gamma>0$, consistent with the second term being subleading in the low temperature expansion.

Considering the field theory as a thermodynamical system and imposing local stability demands that \cite{DeWolfe:2010he}\footnote{$\chi$ is the charge susceptibility and $C_\rho$ is the specific heat at constant volume.}
\be\label{cond1}
\chi=\left(\frac{\partial \rho}{\partial \mu}\right)_T>0,
\ee
and
\be\label{cond3}
C_{\rho}=T\left(\frac{\partial s}{\partial T}\right)_{\rho}=-T\left[\frac{\partial^2 f}{\partial T^2}-\left(\frac{\partial^2 f}{\partial T \partial \mu}\right)^2\frac{1}{\frac{\partial^2 f}{\partial \mu^2}}\right]>0.
\ee
Applying these to (\ref{ansatz}) in the $T/\m\rightarrow 0$ limit gives the constraints $\alpha>0$ and $\gamma>1$. 

Examining the speed of sound $v_s$ of our system also allows us to establish a constraint. To ensure causality, we impose \be0\le v_s\le 1,\label{soundspeedreq}\ee with the speed of sound given by \cite{Herzog:2008he}
\be\label{velmut}
v_s^2=-\frac{\colc{\pare{\ppderivative{f}{T}}\rho^2+\pare{\ppderivative{f}{\mu}}s^2-2\pare{\psderivative{f}{T}{\mu}}\rho s}}{(sT+\rho\mu)\colc{\pare{\ppderivative{f}{T}}\pare{\ppderivative{f}{\mu}}-\pare{\psderivative{f}{T}{\mu}}^2}}  ,
\ee
where $\rho$ and $s$ are the charge and entropy densities. For $\gamma>1$, this implies the stronger bound of $\alpha\ge1$. This is the same bound as derived in section \ref{gaugefield} from consideration of the bulk dual of field theories. It is interesting that it arises from very general circumstances in both cases.

\subsubsection*{Zero temperature}

In the zero temperature limit of ansatz (\ref{ansatz}) only the first term survives, so that $f\propto -\mu^{\alpha+1}$. In this case, the only condition for local stability is given by equation (\ref{cond1}), which trivially leads to $\rho\propto\mu^\alpha$ with $\alpha>0$. Computing the speed of sound and enforcing causality leads again to $\a\ge 1$.

\subsubsection*{General conformal theory}

For a conformal field theory in $d$ spacetime dimensions, the most general free energy density is
\be\label{conf.fenergy}
f=- \mu^dg\pare{\frac{T}{\mu}},
\ee
where $g(x)$ is an arbitrary dimensionless function. Local stability depends on the details of the function $g$, and a general statement is not possible at this point. To ensure causality, we compute equation (\ref{velmut}), finding the speed of propagation to be
\be\label{confspeed}
v_s^2=\frac{1}{d-1},
\ee
from which it follows directly that a conformal theory obeys requirement (\ref{soundspeedreq}) only in dimension $d\ge 2$. This result is trivial, as sound waves are not possible if there are no spacelike dimensions to propagate in.

\subsubsection*{Free fermions}
\label{freeferm}

As an example, we will compute the $\rho-\mu$ relationship for a system of free fermions. In the grand canonical ensemble, the partition function for spin $1/2$ particles of charge $q$ in a 3 dimensional box and subjected to a large chemical potential is
\begin{equation}
\mathcal{Z}(\m,T)
	=\prod_{\vec{n}}(1+e^{-\beta(E_{\vec{n}}-\mu q)}),
\end{equation}
where the product is over available momentum levels. The partition function for antiparticles follows with the replacement $q\rightarrow -q$ so we include antiparticles by considering the total partition function $\tilde{\mathcal{Z}}(\m,T)=\mathcal{Z}(\mu,T)\mathcal{Z}(-\mu,T)$. 
Passing to the continuum limit, approximating the fermions as massless, and setting $q=1$, the resultant charge density is
\begin{eqnarray}
	\rho
	=\frac{\mu^3}{3\pi^2}+\frac{\mu T^2}{3}.
\end{eqnarray}
The dominant power in this case is the same as is expected in a generic conformal field theory. 


\section{General holographic field theories at finite density}\label{gaugefield}

It was shown in the previous section how local stability and causality lead to $\a\ge1$. In this section, under mild assumptions, we investigate the Born-Infeld and Maxwell actions in the large $\m$ regime and observe to what extent they fall under the general results from section \ref{bg}.

\subsection{Finite density}

To find constraints on the $\rho-\mu$ relation in holographic field theories, we begin by studying very general systems with the minimal structure of a conserved charge. The holographic dictionary gives that a conserved charge in the field theory is dual to a massless $U(1)$ gauge field $A$ in the bulk \cite{Minwalla:1997ka}. If the gauge field is a function only of the radial coordinate $r$, the chemical potential and the charge density are encoded in the behaviour of $A$ as
\be\label{mu_inf}
\m=A_t(\infty)
\ee
and
\be
\rho=-\frac{\partial S_{E}}{\partial A_{t}(\infty)},
\ee
where $S_E$ is the Euclidean action evaluated on the saddle-point and the derivative is taken holding other sources fixed. As discussed in \cite{Kobayashi:2006sb}, an equivalent expression for the charge density is\footnote{Generically, $A_t$ is a cyclic variable, so that the conjugate momentum is conserved, and we may evaluate this expression at any $r$.}
\be\label{density}
\rho=\left( \frac{1}{d-2} \right)\frac{\d\mathcal{L}}{\d(\d_r A_t)},
\ee
where the normalization of $\rho$ has been chosen for later convenience. After writing down the gravitational lagrangian, our prescription for computing the charge density at a given chemical potential is to solve the equations of motion with a fixed boundary condition for the gauge field, equation (\ref{mu_inf}), before reading off the density using equation (\ref{density}).

\subsection{Gauge field actions}

To include a gauge field in our AdS/CFT construction, we simply include it in the bulk action. Two gauge field lagrangians that have appeared in holographic studies are the Maxwell and the Born-Infeld lagrangians. Typically, the Maxwell action is used in bottom-up holographic models while the Born-Infeld action appears in the study of brane dynamics. Below, in section \ref{probes} we will consider holographic models using both types of lagrangians. However, much insight can be gained by investigating these actions under generic conditions. Therefore, in this section, we study general versions of these two lagrangians, at fixed temperature and large chemical potential, in the probe approximation.\footnote{In the probe approximation, we assume there is no backreaction on the gravity metric. This is enforced in this case by studying the gauge field lagrangian on a fixed background geometry.} Interpreting our results using (\ref{mu_inf}) and (\ref{density}), we will develop some constraints for the $\rho-\mu$ relationship in holographic theories described by these actions.

\subsubsection*{The Maxwell action}\label{generalym}

Consider a gauge field described by the Maxwell action $\int\sqrt{-g}F^2$ in a general background of the form
\be
ds^2=g^{FT}_{\m\nu}(r)dx^\m dx^\nu+g_{rr}(r)dr^2.
\ee
If we assume homogeneity in the field theory directions and consider a purely electrical gauge field (keeping only its time-component), the lagrangian is simply
\be
\mathcal{L}=g(r)\pare{\partial_r A_t}^2,
\ee
for some function $g(r)$. From this we find 
\be\label{rho_ym}
\r=
\left(\frac{2}{d-2}\right) g(r)\d_rA_t.
\ee

In the systems considered below, the spacetime either has a horizon or smoothly cuts off at some radius $r_\textrm{min}$. The value of the gauge field at this point is a boundary condition for the problem. Below, $A_t(r_\textrm{min})$ is either zero or a constant, neither of which affect the $\rho-\mu$ behaviour; we take $A_t(r_\textrm{min})=0$ here. Integrating (\ref{rho_ym}), we find
\be
\mu=
\rho\left(\frac{d-2}{2}\right) \int_{r_\textrm{min}}^\infty \frac{dr}{g(r)}.
\ee
Provided the integral is finite, we have
\be\label{ym}
\rho\propto\mu.
\ee
Thus, for any holographic field theory with the gauge field described only by the Maxwell lagrangian in a fixed metric we have $\alpha=1$.

\subsubsection*{The Born-Infeld action}\label{generalbia}

The Born-Infeld action is the non-linear generalization of Maxwell electrodynamics and is the appropriate language in which to describe the dynamics of gauge fields living on branes. Assuming homogeneity in the field theory directions, so that the gauge potential varies only with the radial direction, these systems are governed by an action of the generic form\footnote{$g(r)$ and $h(r)$ are arbitrary functions; $g(r)$ is not related to the previous discussion.}

\begin{eqnarray}\label{bi_gen}
	\mathcal{L}=\sqrt{g(r)-h(r)(\partial_rA_t)^2},
\end{eqnarray}
where again, we take $A_t$ to be the only non-zero part of the gauge field. The charge density is given by the constant of motion
\begin{eqnarray}
	\rho=\left(\frac{1}{d-2}\right)\frac{h( r) \d_rA_t(r )}{\sqrt{g(r)-h(r)(\partial_rA_t)^2}}.
\end{eqnarray}
Here, we assume that the gauge field is sourced by a charged black hole horizon at $r_+$.\footnote{To have a non-trivial field configuration, a source for the gauge field in the bulk is required. In the low temperature, horizon-free versions of these models, this source is given by lower dimensional branes wrapped in directions transverse to the probe branes \cite{Witten:1998xy}. } Euclidean regularity of the potential $A_t$ fixes its value at the horizon as $A_t(r_+)=0$ \cite{Kobayashi:2006sb}. Then, we can integrate to find
\begin{eqnarray}\label{bi_mu}
	\mu=\int_{r_+}^\infty dr\sqrt{\frac{g(r)}{h(r)}}\frac{(d-2)\rho}{\sqrt{h(r)+(d-2)^2\rho^2}}.
\end{eqnarray}

To extract the large $\rho$ behaviour, we split the integral at $\Lambda\gg1$. For $\rho\gg\Lambda$, the integral from $r_+$ to $\Lambda$ approaches a constant, while the functions in the integral from $\Lambda$ to $\infty$ can be approximated by their large $r$ forms, which will be denoted with a $\infty$ subscript. The expression for the chemical potential now becomes
\begin{eqnarray}
	\mu\approx\int_{r_+}^\Lambda dr\sqrt{\frac{g(r)}{h(r)}}+\int_{\Lambda}^\infty dr\sqrt{\frac{g_\infty(r)}{h_\infty(r)}}\frac{(d-2)\rho}{\sqrt{h_\infty(r)+(d-2)^2\rho^2}}.
\end{eqnarray}
The $\rho$ dependence of $\mu$ comes from the second term. If $g_\infty(r)/h_\infty(r)\approx r^{2m}$ and $h_\infty(r)\approx r^n$, by putting $x=r/\rho^{2/n}$ we find that
\begin{eqnarray}\label{BImurho}
	\mu\sim\rho^{(2+2m)/n}\int_{\frac{r_+}{\rho^{2/n}}}^\infty dx \frac{x^m}{\sqrt{x^n+1}}.
\end{eqnarray}
The convergence of the integral here requires that $n/(2+2m)>1$,
resulting in the relationship
\be
\rho\propto\mu^{\alpha}\quad\mbox{with}\quad\alpha>1,
\ee
where the power $\alpha$ depends on the specific bulk geometry.



\section{Holographic probes}\label{probes}

With the general constraints of the previous sections in hand, we move on to study particular holographic field theories in the probe approximation, to see which specific values of $\alpha$ are realized. Here, we study two common probe configurations that have arisen in previous holographic studies. These are extensions of the actions considered in section \ref{gaugefield}. First, we examine probe branes in the black brane background using the Born-Infeld action. Then, we move on to the phenomenological perspective, in which we write down an effective gravity action without appealing to the higher dimensional string theory. In this approximation, using the Maxwell action, we look at the gauge field in both the planar Schwarzschild black hole and soliton backgrounds, with and without a coupling to a scalar field.

In both cases, in the systems we consider, the only sources in the field theory are the temperature $T$ and chemical potential $\mu$. Below, we fix $T$ and work at large $\mu$ (such that $\mu/T\gg1$). In this regime, we look for a relationship $\rho\propto\mu^\alpha+\dots$, where the dots denote terms higher order in $T/\mu$. 

\subsection{Probe branes and the Born-Infeld action}\label{brane.section}

In the systems we will consider here, the background consists of $N_c$ D-branes; in the large $N_c$ limit, these branes are replaced with a classical gravity metric. In this regime, fundamental matter is added by placing $N_f$ probe branes in the geometry \cite{Karch:2002sh}.

\subsubsection*{The brane action}

Assuming that the background spacetime metric $G_{\mu\nu}$ is given, the action governing the dynamics of a single D$q$ probe brane is the Born-Infeld action
\begin{eqnarray}
	S\propto\int d^{q+1}\sigma e^{-\phi}\sqrt{-\textrm{det}(g_{ab}+2\pi \a^{\prime}F_{ab})} .
\end{eqnarray}
Here, latin indices refer to brane coordinates and greek indices denote spacetime coordinates, while $X^\mu(\sigma^a)$ describes the brane embedding. $g_{ab}$ is the induced metric on the probe brane given by $g_{ab}=\partial_a X^\mu \partial_bX^\nu G_{\mu\nu}$,
$F_{ab}$ is the field strength for the $U(1)$ gauge field on the brane, and $\phi$ is the dilaton field. Following the previous discussion, the only component of the gauge field we choose to turn on is $A_t$, additionally, we assume it depends only on the radial coordinate $r$, $A_t=A_t(r)$. Considering that the probe brane is extended in the $r$ direction and the spacetime metric is diagonal, the lagrangian simplifies to
\begin{eqnarray}
	\mathcal{L}\propto e^{-\phi}\sqrt{-\textrm{det}(g_{ab})\left(1+\frac{(\partial_rA_t)^2}{g_{tt}g_{rr}}\right)} ,
\end{eqnarray}
where we rescaled $A_t$ to absorb the $2\pi \a^{\prime}$ term. In the notation of equation (\ref{bi_gen}), we can write
\begin{eqnarray}
	g(r)&=&-\textrm{det}(g_{ab})e^{-2\phi} , \label{borninfeldg}\\
	h(r)&=&\frac{\textrm{det}(g_{ab})e^{-2\phi}}{g_{tt}g_{rr}} \label{borninfeldh}.
\end{eqnarray}

\subsubsection*{The background}

For $N_c$ D$p$ branes, at large $N_c$, the high temperature background is the black D$p$ brane metric, given by\footnote{More details on this solution can be found in \cite{Mateos:2006nu}.}
\be
ds^2=H^{-1/2}(-fdt^2+d\vect{x}^2_p)+H^{1/2}\left(\frac{dr^2}{f}+r^2d\Omega^2_{8-p}\right) ,
\ee
with
\begin{equation}
H(r)=\left(\frac{L}{r}\right)^{7-p}, \quad
f(r)=1-\left(\frac{r_+}{r}\right)^{7-p},\quad
e^\phi=g_sH^{(3-p)/4}.
\end{equation}  
$L$ is the characteristic length of the space, while $g_s$ is the string coupling. This metric has a horizon at $r=r_+$.

Our probe D$q$ brane is fixed to share $d-1$ spacelike directions with the D$p$ branes. If $p>d-1$, the fundamental matter propagates on a $d$ dimensional defect and we may consider the extra $p-(d-1)$ directions along the background brane to be compactified, giving an effective $d$ dimensional gauge theory at low energies. Alternatively, we can build a $d-1$ dimensional gauge theory by compactifying one or more of the directions shared by the probe and background branes. Below, we will study field theories that are effectively $3+1$ dimensional using both methods.

We stipulate that the D$q$ probe brane wraps an $S^n$ inside the $S^{8-p}$ and extends along the radial direction $r$. These quantities are related by $q=d+n$. The induced metric on the D$q$ brane is
\be
	ds^2=H^{-1/2}(-fdt^2+d\vect{x}^2_{d-1})+\left(\eta(r)+\frac{H^{1/2}}{f} \right)dr^2+H^{1/2}r^2d\Omega^2_n ,
\ee
where
\begin{equation}
	\eta(r)=\partial_rX^\mu\partial_rX^\nu G_{\mu\nu}-G_{rr} .
\end{equation}
Calculating equations (\ref{borninfeldg}) and (\ref{borninfeldh}) gives\footnote{We leave the constant factors of $g_s$ from $e^\phi$ out of the lagrangian, as our goal here is just the power law dependence.}
\begin{eqnarray}
	g(r)&=&r^{2n}fH^{\frac{1}{2}(p+n-d-3)}\left(\eta(r)+\frac{H^{1/2}}{f} \right) , \\
	h(r)&=&r^{2n}H^{\frac{1}{2}(p+n-d-2)} ,
\end{eqnarray}
from which (\ref{bi_mu}) gives the chemical potential
\begin{eqnarray}
	\mu=\int_{r_+}^\infty dr\frac{(d-2)\rho}{\sqrt{r^{2n}\left( \frac{L}{r} \right)^{(\frac{7-p}{2})(p+n-d-2)}+(d-2)^2\rho^2}}\sqrt{\frac{f\eta(r)}{H^{1/2}}+1} .
	\label{chempot}
\end{eqnarray}

Now, $\eta(r)$ will be some combination of $(\partial_r\chi_i)^2$, where the $\chi_i$ denote the directions of transverse brane fluctuations. By writing down the equations of motion we can observe that $\partial_r\chi_i=0$ is a solution, in which case the probe brane goes straight into the black hole along the radial direction $r$. This describes the high temperature, deconfined regime; we set $\eta(r)=0$ in the following.

For large $\rho$ we find
\begin{eqnarray}
	\rho\propto\mu^{\frac{1}{4}[(p-7)(p-d-2)+(p-3)(q-d)]},
	\label{mu}
\end{eqnarray}
so that for the probe brane systems,
\be
\alpha = \frac{1}{4}[(p-7)(p-d-2)+(p-3)(q-d)].
\label{alphaDq}
\ee
As above, $\alpha$ is constrained as $\alpha>1$ for convergence of the integral. If $\alpha\le1$, the contribution of the constant charge density to the total energy diverges, signalling a breakdown of the probe approximation. At this point, we can use equations (\ref{mu}) and (\ref{alphaDq}) to investigate what type of $\rho-\mu$ behaviours can arise from D$p$-D$q$ brane constructions. 

\subsubsection*{Example: the Sakai-Sugimoto model}

The well-known Sakai-Sugimoto model \cite{Sakai:2004cn} consists of $N_f$ probe D8-$\overline{\textrm{D}8}$ branes in a background of $N_c$ D4 branes compactified on a circle. We have $p=4$, $q=8$, and $d=4$. Putting these numbers into (\ref{mu}) yields
\be
	\rho\propto\mu^{5/2},
\ee
consistent with previous results \cite{Rozali:2007rx,Bergman:2007wp}.

\subsubsection*{$\rho-\mu$ in $3+1$ dimensional probe brane theories}

Equation (\ref{alphaDq}) determines the dominant power law behaviour in all D$p$-D$q$ configurations relevant to $3+1$ dimensional field theory. As discussed above, we can set the number of shared probe and background directions to be $d-1=3$ or put $d-1=4$ and demand one of the the spacelike shared directions to be periodic; see table \ref{summary1} for the results. The power $\alpha=3$ is an upper bound for the $3+1$ dimensional probe brane gauge theories we have considered.

Our calculation above involves only the Born-Infeld action for the probe brane and in particular neglects any possible Chern-Simons terms that appear due to the coupling between the brane and a spacetime tensor field. The Chern-Simons term is important in the D4-D4 system, for example \cite{VanRaamsdonk:2009gh}.

\subsection{Bottom-up models and the Einstein-Maxwell action}
\label{phen_probe}

We now turn our attention to bottom-up AdS/CFT models in the probe regime. To construct a phenomenological gauge / gravity model, we begin with a theory of gravity with a cosmological constant, such that the geometry is asymptotically AdS. To study the relationship between charge density and chemical potential in the dual field theory, we demand that there must be a gauge field in the bulk. At this point, our model has the ingredients for us to compute our desired result. But, one may ask what type of extensions are possible. Motivated by superconductivity and superfluidity studies, we will consider also a charged scalar field in our gravity theory. Adding a scalar field alters the dynamics of the system, notably resulting in different phases \cite{Klebanov:1999tb,Gubser:2008px}. When the scalar field takes on a non-zero expectation value, this breaks the $U(1)$ gauge symmetry in the bulk and corresponds to the presence of a $U(1)$ condensate in the boundary theory.

The particular model we study is the Einstein-Maxwell system with a charged scalar field:
\begin{eqnarray}
	S=\int d^{d+1}x\sqrt{-g}\left\{\mathcal{R}+\frac{d(d-1)}{L^2}-\frac{1}{4}F^{\mu\nu}F_{\mu\nu}-|\d_\mu\psi-iqA_\mu\psi|^2-V(|\psi|)\right\} .
	\label{totalaction1}
\end{eqnarray}
Different dual field theories may be obtained by considering this action in different regimes and with different parameters. Below, we make the following ansatz for the gauge and scalar fields:
\be
A=\phi(r)dt, \:\:\:   \psi=\psi(r).
\ee
The $r$ component of Maxwell's equations will give that the phase of the complex field $\psi$ is constant, so without loss of generality we take $\psi$ real. For the remainder of the study, we choose units such that $L=1$ and consider the potential $V(\psi)=m^2 \psi^2$.

\subsubsection*{The probe limit}

To get the probe approximation for the system described by (\ref{totalaction1}), we rescale $\psi\rightarrow\psi/q$ and $A\rightarrow A/q$ before taking $q\rightarrow\infty$ while keeping the product $q\mu$ fixed (to maintain the same $A-\psi$ coupling). The gauge and scalar fields decouple from the Einstein equations and we study the fields in a fixed gravitational background.

The background is governed by the action
\be
S=\int d^{d+1}x\sqrt{-g}\left\{\mathcal{R}+d(d-1)\right\} .
\label{backgroundaction}
\ee
One solution here is the planar Schwarzschild-AdS black hole, given by
\be\label{schwarzschildAdS}
	ds_{bh}^2=(-f_{bh}(r)dt^2+r^2dx_idx^i)+\frac{dr^2}{f_{bh}(r)},
\ee
with
\begin{equation}\label{f-schwarzschildAdS}
	f_{bh}(r)=r^2\left( 1-\frac{r_+^d}{r^d}  \right) ,
\end{equation}
where $r_+$ is the black hole horizon. Below, we consider two systems in the Schwarzschild-AdS background: the probe gauge field, and the probe gauge and scalar fields.

\subsubsection*{Computing $\mu$ and $\rho$}

If the kinetic term for the gauge theory on the gravity side is the Maxwell lagrangian,
\be \label{maxwell}
\mathcal{L}=\frac{1}{4}\sqrt{-g}F_{\m\n}F^{\m\n} ,
\ee
then for an asymptotically AdS space the field equation for the time component of the gauge field is 
\be
\phi''+\frac{d-1}{r}\phi'+\dots=0 ,
\ee
where $'$ denotes an $r$ derivative and $\dots$ denotes terms that have higher powers of $1/r$. The solution is
\be \label{phi_sol}
\phi(r)=\phi_1+\frac{\phi_2}{r^{d-2}}+\dots.
\ee
Recalling that $\phi(\infty)=\mu$ determines that $\phi_1=\mu$, while we can plug (\ref{phi_sol}) into (\ref{maxwell}) and compute, using (\ref{density}), that $\phi_2=\rho$. We have that
\be\label{phi_expand}
\phi(r) = \mu- \frac{\rho}{r^{d-2}} +\dots,
\ee
so that in practice, below, we just have to read off the coefficients of the leading and next to leading power of $1/r$ to find the chemical potential and the charge density.

\subsubsection*{The scalar field}
Solving the scalar field equation at large $r$ in an asymptotically AdS space results in the behaviour
\be
\psi=\frac{\psi_1}{r^{\lambda_-}}+\frac{\psi_2}{r^{\lambda_+}}+\dots ,
\ee
where
\be\label{bound.op.dim}
	\lambda_\pm=\frac{1}{2}\left\{ d\pm\sqrt{d^2+4m^2} \right\}.
\ee
For $m^2$ near the Breitenlohner-Freedman (BF) bound \cite{Breitenlohner:1982jf,Breitenlohner:1982bm}, in the range $-(d-1)^2/4\ge m^2\ge-d^2/4$, the choice of either $\psi_1=0$ or $\psi_2=0$ results in a normalizable solution \cite{Klebanov:1999tb}. For $m^2>-(d-1)^2/4$, $\psi_1$ is a non-normalizable mode and $\psi_2$ is a normalizable mode. For the cases with the scalar field, we define our field theory by taking $\psi_1=0$, so that we never introduce a source for the operator dual to the scalar field.

\subsubsection{The probe gauge field}

Here, we study the probe gauge field, without the scalar field, in the Schwarzschild-AdS background (\ref{schwarzschildAdS}). The equation of motion for $\phi$ is
\begin{eqnarray}
	\phi''+\frac{d-1}{r}\phi'=0 .
	\label{phi_probe_equation}
\end{eqnarray}

Regularity at the horizon demands that $\phi(r_+)=0$ and the AdS/CFT dictionary gives $\phi(\infty)=\mu$, leading to
\begin{eqnarray}
	\phi(r)=\mu\left(1 -\frac{r_+^{d-2}}{r^{d-2}} \right) .
\end{eqnarray}
Then, applying (\ref{phi_expand}), we have
\be\label{rhoprobegauge}
\r=\m r_+^{d-2}. 
\ee
The horizon $r_+$ depends only on the temperature, $T=r_+d/4\pi$,\footnote{For a Euclidean metric $ds^2=\alpha(r)d\tau^2+\frac{dr^2}{\beta(r)}$ with periodic $\tau=it$ coordinate and $\alpha(r_+)=\beta(r_+)=0$, regularity at the horizon demands that the temperature (the inverse period of $\tau$) be given by $T=\sqrt{\alpha'(r_+)\beta'(r_+)}/4\pi$.} so this is a linear relationship between $\rho$ and $\mu$, in accordance with (\ref{ym}).

\subsubsection{Adding a scalar field}
\label{probesect}

We now turn on the scalar field in (\ref{totalaction1}), and consider the dynamics in the Schwarzschild-AdS background (\ref{schwarzschildAdS}).

The field equations become
\be
	\psi''+\left( \frac{f_{bh}'}{f_{bh}} +\frac{d-1}{r} \right)\psi'+\left(\frac{q^2\phi^2}{f_{bh}^2}-\frac{m^2}{f_{bh}}\right)\psi=0 \label{bhpsiprobe} ,
\ee
\be
	\phi''+\frac{d-1}{r}\phi'-\frac{2q^2\psi^2}{f_{bh}}\phi=0 . \label{bhphiprobe}
\ee 
At this point, we can scale $q$ to 1 by scaling $\phi$ and $\psi$, and so $m$ is the only parameter here. 

The coupling allows the gauge field to act as a negative mass for the scalar field. At small chemical potentials, $\psi=0$ is the solution. As we increase $\m$, the effect of the gauge field on the scalar field becomes large enough such that the effective mass of the scalar field drops below the BF bound of the near horizon limit of the geometry, so that a non-zero profile for $\psi$ is possible, and we have a phase transition to the field theory state with broken $U(1)$ symmetry. A smaller (more negative) squared mass results in a smaller critical chemical potential, at which the scalar field turns on.

\begin{figure}[htb!]
\centering%
\includegraphics[scale=1.5]{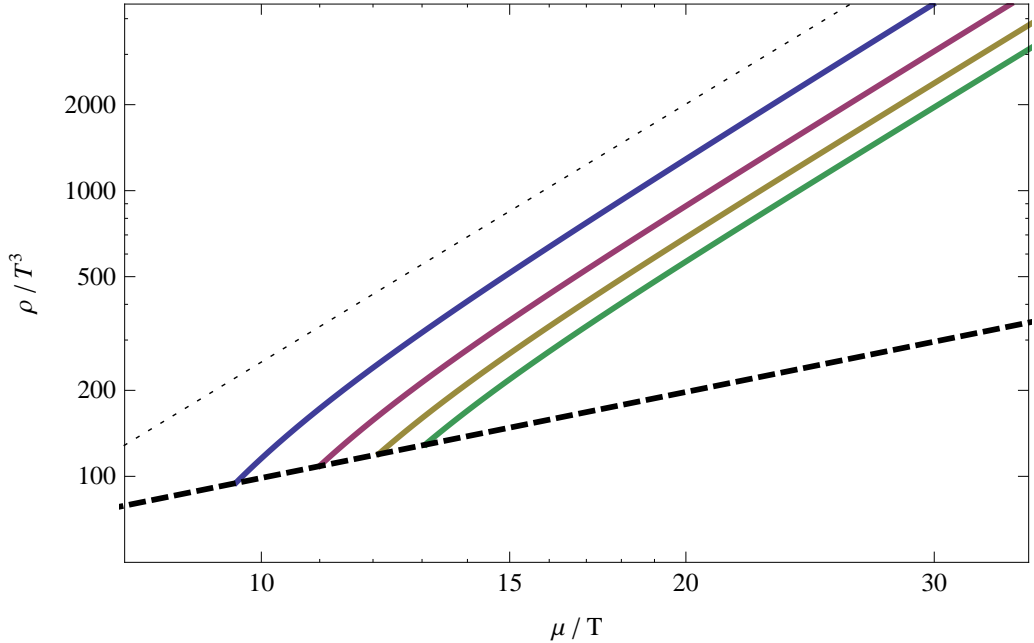}
\caption{Charge density versus chemical potential for the probe gauge and scalar fields, section \ref{probesect}, on a log-log scale. The thick dashed line is for the system with no scalar field for which, analytically, $\rho\propto\mu$. At a critical chemical potential, depending on the mass of the scalar field, configurations with non-zero scalar field become available. The thin dotted line is a model power law $\rho\propto\mu^3$, as described in equation (\ref{cprobe}). From left to right, the thick solid lines are for scalar field masses $m^2=-15/4$, $-14/4$, $-13/4$, and $-3$. A more negative scalar field mass results in a denser field theory state at a given chemical potential.}
\label{5d_probe_bh}
\end{figure}

Using a simple shooting method, for $d=4$ we numerically solve equations (\ref{bhpsiprobe},  \ref{bhphiprobe}) and arrive at the relationship
\begin{eqnarray}\label{cprobe}
	\rho=c^p_{bh}(m^2)\mu^3 ,
\end{eqnarray}
where $c^p_{bh}(m^2)$ is a scaling constant that depends on the mass of the scalar field. The coupling to the scalar field has resulted in the larger power ($\alpha=3$) in the scaling of $\rho$. A smaller squared mass corresponds to a larger value of $c^p_{bh}$ and, for a given chemical potential, is dual to field theory with a higher charge density. In figure \ref{5d_probe_bh}, we can see the existence of a denser state when the scalar field turns on as well as the relative relation between the mass of the scalar field and the charge density in the field theory.

\subsubsection{The soliton probe}
\label{solprobe}

Motivated by recent work \cite{Nishioka:2009zj,Horowitz:2010jq,Basu:2011yg}, we now add more structure to the bulk theory in the form of an extra periodic dimension. To model a $3+1$ dimensional field theory, we set $d=5$ and stipulate that this includes one periodic spacelike coordinate $w$ of length $2\pi R$. At energies much less than the scale set by this length, $E\ll 1/R$, the dual field theory will be effectively $3+1$ dimensional. The extra dimension sets another scale for the field theory and enables a richer phase structure in the system.\footnote{The phase diagram including both black hole and soliton solutions, was studied in \cite{Horowitz:2010jq} for a $2+1$ dimensional field theory in the context of holographic superconductors and in \cite{Basu:2011yg} for a $3+1$ dimensional field theory in the context of holographic QCD and colour superconductivity.}

With the extra periodic direction, there is another solution to the background described by (\ref{backgroundaction}). This is the AdS-soliton, given as the double-analytic continuation of the Schwarzschild-AdS solution (\ref{schwarzschildAdS}):
\be\label{solitonAdS}
	ds_{sol}^2=(r^2dx_\m dx^\m+f_{sol}(r)dw^2)+\frac{dr^2}{f_{sol}(r)} ,
\ee
with
\be
	f_{sol}=r^2\left( 1-\frac{r_0^5}{r^5}  \right) .
\ee
Here, $r_0$ is the location of the tip of the soliton. For regularity, it is fixed by the length of the $w$ dimension as
\begin{eqnarray}
	r_0=\frac{2}{5R} .
\end{eqnarray}
By computing the free energy of the systems, it can be shown that the soliton background dominates over the black hole background for small enough temperatures and chemical potentials. As the temperature or chemical potential is increased, there is a first order phase transition to the black hole, which is the holographic version of a confinement / deconfinement transition.

For zero scalar field, the soliton can be considered at any temperature and chemical potential; the period of the Euclidean time direction defines the temperature while $\phi=\m=\textrm{constant}$ is a solution to the field equations. In this case, $\rho=0$ and we do not have an interesting $\r-\m$ relation. Considering a non-zero scalar field provides a source for the gauge field and allows non-trivial configurations. 

In the soliton background (\ref{solitonAdS}),  the equations of motion are
\be
	\psi''+\left( \frac{f_{sol}'}{f_{sol}}+\frac{4}{r} \right)\psi'+\left(\frac{q^2\phi^2}{r^2f_{sol}}-\frac{m^2}{f_{sol}}\right)\psi=0 ,
\ee
\be
	\phi''+\left( \frac{f_{sol}'}{f_{sol}}+\frac{2}{r} \right)\phi'-\frac{2q^2\psi^2}{f_{sol}}\phi=0 .
\ee
As in the black hole case, at this point we can set $q=1$ by scaling the fields. 

After numerically integrating, we have
\begin{eqnarray}\label{solpowerlaw}
	\rho=c^p_{sol}(m^2)\mu^4.
\end{eqnarray}
Compared to the black hole case, above, we find a larger power of $\mu$. At large densities, the average distance between charges becomes small compared to the size $R$ of the periodic direction. In this limit, the system becomes effectively higher dimensional and so we would expect a larger power $\alpha$ in the $\rho-\mu$ relationship. The numerics were consistent with this result.

\begin{figure}[htb!]
\centering%
\includegraphics[scale=1.5]{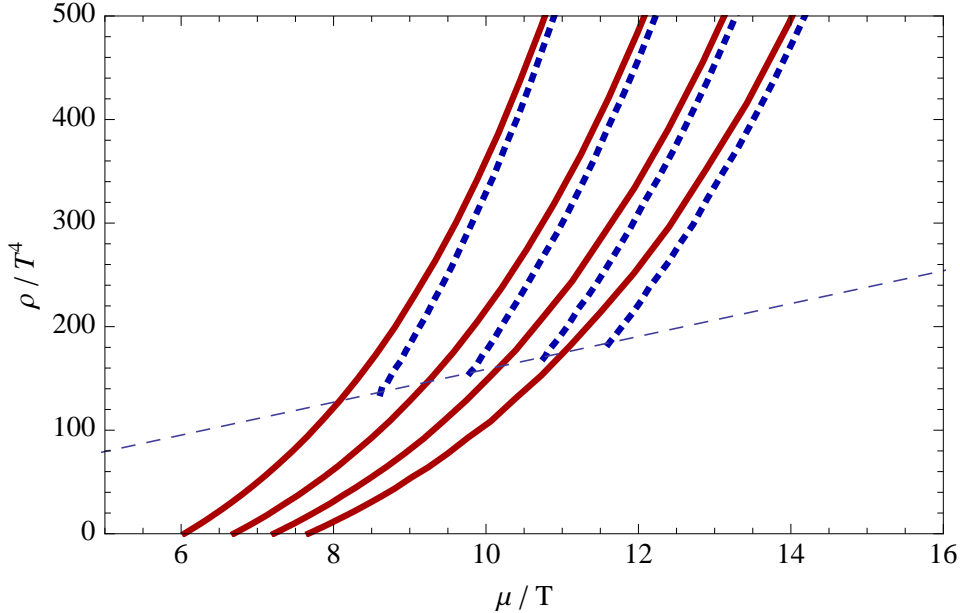}
\caption{Charge density versus chemical potential for the probe gauge and scalar fields in the soliton background, section \ref{solprobe}, and the $d=5$ black hole background, section \ref{probesect}. The thin dashed line is the probe gauge field in the black hole background for which, analytically, $\rho\propto\mu$. The thick solid lines are the soliton results (from left to right, the squared mass of the scalar field is $-22/4$, $-5$, $-18/4$, and $-4$) while the thick dashed lines are the black hole results (again, from left to right, $m^2=-22/4$, $-5$, $-18/4$, and $-4$). Each of the thick lines approaches the power law $\rho\propto\mu^4$, equation (\ref{solpowerlaw}). At a given chemical potential, the soliton background gives a field theory in a denser state.}\label{solbhcompa}
\label{6d_probe_sol}
\end{figure}

As can be seen in figure \ref{6d_probe_sol}, a more negative mass squared results in a smaller critical chemical potential and a denser field theory state at a given chemical potential. This is as expected by comparing the structure of the equations to those in the black hole case. Further, at a given chemical potential, the soliton solution corresponds to a denser field theory state than the black hole solution with the same scalar field mass.

\section{$\rho-\mu$ in backreacted systems}\label{br}

Despite our analysis in section 2 relying on the probe approximation, it is interesting to ask how much of a difference allowing for backreaction on the bottom-up models could make to the $\r-\m$ relation and the bounds found previously. Henceforth we generalize the bottom-up model introduced in section \ref{phen_probe} and allow for the backreaction of the gauge and scalar field on the metric. Recall that the action is
\begin{eqnarray}
	S=\int d^{d+1}x\sqrt{-g}\left\{\mathcal{R}+d(d-1)-\frac{1}{4}F^{\mu\nu}F_{\mu\nu}-|\d_\mu\psi-iqA_\m\psi|^2-m^2 \psi^2\right\} .
	\label{totalaction}
\end{eqnarray}
We start by studying the well-known Reissner-Nordstrom-AdS (RN-AdS) solution to the Einstein equation, in which $\psi=0$. Later, we allow the scalar field to acquire a non-zero profile and investigate its consequences on the $\r-\m$ profile. We finish with the investigation of the backreacted version of the solitonic solution.

\subsection{Charged black holes}

The backreacted solution with no scalar field is the planar RN-AdS black hole, given by

\be
	ds^2=(-f_\textrm{RN}(r)dt^2+r^2dx_idx^i)+\frac{dr^2}{f_\textrm{RN}(r)} ,
\ee
with\footnote{We parametrize this solution in terms of the location of the horizon $r_+$ and the asymptotic value of the gauge field (the chemical potential $\mu$) instead of the usual choices of the charge and mass of the black hole.}
\be
f_\textrm{RN}(r)=r^2\left(1- \left(1+\frac{(d-2)\m^2}{2(d-1)r_+^2}\right)\frac{r_+^d}{r^d} + \frac{(d-2)\m^2}{2(d-1)}\frac{r_+^{2(d-2)}}{r^{2(d-1)}}  \right) .
\ee
The gauge potential is
\be
	\phi(r)=\mu\left( 1-\frac{r_+^{d-2}}{r^{d-2}} \right) ,
\ee
so that, using (\ref{phi_expand}), we have $\r= \m r_+^{d-2}$. Here, the horizon $r_+$ can be expressed as a function of the temperature and chemical potential through the Hawking temperature
\be
	T=\frac{1}{4\pi}\left( dr_+-\frac{(d-2)^2\m^2}{2(d-1)r_+} \right) .
	\label{T_rn}
\ee

Eliminating $r_+$ in favour of $\rho$ and $\mu$ in (\ref{T_rn}), we may solve for $\r$ to find
\be
\r=\left( \frac{(d-2)^2}{2d(d-1)} \right)^{\frac{d-2}{2}}\m^{d-1}\left[ \left(\frac{2(d-1)}{d}\right)^{\frac{1}{2}}\frac{2\pi T}{(d-2)\m} + \sqrt{ 1+\frac{8\pi^2(d-1)T^2}{d(d-2)^2\m^2} } \right]^{d-2} .
\label{rhoRN}
\ee
Notice that the dominant power in the $\rho-\mu$ relationship is $\mu^{d-1}$, as expected in a $d$ dimensional conformal field theory. For $d=4$, the particular large $\mu$ expansion is
\be\label{d4_RN}
\r=\frac{1}{6}\m^3+\frac{\pi }{\sqrt{6}}\m^2T+\frac{1}{2} \pi ^2 \mu  T^2+\frac{1}{4} \sqrt{\frac{3}{2}} \pi ^3 T^3+\dots .
\ee

\subsection{Hairy black holes}\label{hairy}

If we turn on the scalar field, an analytic solution to the equations of motion is no longer possible and we turn to numerical calculation. We take as our metric ansatz
\begin{eqnarray}
ds^2=-g(r)e^{-\chi(r)}dt^2+\frac{dr^2}{g(r)}+r^2(dx_idx^i) ,
\end{eqnarray}
where $g(r)$ will be fixed to have a zero at $r_+$, giving a horizon. We arrive at the following equations of motion:
\be
\psi''+\left( \frac{g'}{g}-\frac{\chi'}{2}+\frac{d-1}{r} \right)\psi'+\frac{1}{g}\left(\frac{q^2\phi^2e^\chi}{g}-m^2\right)\psi=0 \label{eom1}, 
\ee
\be
\phi''+\left( \frac{\chi'}{2}+\frac{d-1}{r} \right)\phi'-\frac{2q^2\psi^2}{g}\phi=0 \label{eom2} ,
\ee
\be
\chi'+\frac{2r\psi'^2}{d-1}+\frac{2rq^2\phi^2\psi^2e^\chi}{(d-1)g^2}=0 \label{eom3} ,
\ee
\be
g'+\left( \frac{d-2}{r}-\frac{\chi'}{2} \right)g+\frac{re^\chi \phi'^2}{2(d-1)}+\frac{rm^2\psi^2}{d-1}-dr=0.
\label{eom4}
\ee
The first two equations can be derived via the Euler-Lagrange equations for $\phi$ and $\psi$, while the final two equations are the $tt$ and $rr$ components of Einstein's equation.

In this system, as in the probe case, section \ref{probesect}, at small chemical potentials the scalar field is identically zero. As we increase the chemical potential above a critical value, the system undergoes a second order phase transition to a state with non-zero scalar field. When the scalar field condenses, the corresponding field theory is in a denser state at the same chemical potential than for the system without scalar field.

We solve the equations numerically for $d=4$, to yield the result, in the phase with the scalar field,
\bea\label{bhpowerlaw}
\r=c_{bh}(q,m^2)\m^3.
\eea
As we increase the charge or decrease the mass squared of the scalar field, the critical chemical potential, at which the scalar condenses, decreases, while the scaling coefficient $c_{bh}$ increases. The scaling coefficient $c_{bh}(q,m^2)$ is, in all cases we checked, larger than the coefficient of the $\mu^3$ term in the AdS-Reissner-Nordstrom black hole, equation (\ref{d4_RN}), indicating that the density scales faster with the chemical potential when the scalar field is present. 

When we include metric backreaction for the black hole, the dominant power in the $\rho-\mu$ relationship is greater than the probe case when there is no scalar field and is the same as the probe case when there is a scalar field, indicating that, at least for the systems considered, the bounds found for the $\r-\m$ behaviour apply to the backreacted cases as well.

\subsection{Backreacted soliton}

Motivated by the form of the soliton background (\ref{solitonAdS}) we choose the metric ansatz
\begin{equation}
ds^2=\frac{dr^2}{r^2B(r)}+r^2\left(e^{A(r)}B(r)dw^2-e^{C(r)}dt^2+dx_{i}dx^{i}\right) ,
\end{equation}
where we constrain $B(r_0)=0$ so that the tip of the soliton is at $r_0$. The field and Einstein equations give
\begin{equation}
\psi''+\left(\frac{6}{r}+\frac{A'}{2}+\frac{B'}{B}+\frac{C'}{2}\right)\psi'+\frac{1}{r^2B}\left(\frac{e^{-C}(q \phi)^2}{r^2}-m^2\right)\psi=0,\label{soleom1}
\end{equation}
\begin{equation}
\phi''+\left(\frac{4}{r}+\frac{A'}{2}+\frac{B'}{B}-\frac{C'}{2}\right)\phi'-\frac{2\psi^2q^2 \phi}{r^2 B}=0,\label{soleom2}
\end{equation}
\begin{multline}
B'\left(\frac{4}{r}-\frac{C'}{2}\right)+B\left(\psi'^2-{1 \over 2} A'C'+\frac{e^{-C}\phi'^2}{2r^2}+\frac{20}{r^2}\right)+\\
+\frac{1}{r^2}\left(\frac{e^{-C}(q\phi)^2\psi^2}{r^2}+m^2\psi^2-20\right)=0,\label{soleom5}
\end{multline}
\begin{equation}
C''+{1 \over 2} C'^2+\left(\frac{6}{r}+\frac{A'}{2}+\frac{B'}{B}\right)C'-\left(\phi'^2+\frac{2(q\phi)^2\psi^2}{r^2B}\right)\frac{e^{-C}}{r^2}=0 ,\label{soleom4}
\end{equation}
\begin{equation}
A'=\frac{2r^2C''+r^2C'^2+4rC'+4r^2\psi'^2-2e^{-C}\phi'^2}{r(8+rC')}.\label{soleom3}
\end{equation}

We solve equations (\ref{soleom1}-\ref{soleom4}) numerically with asymptotically AdS boundary conditions before integrating (\ref{soleom3}) to find $A$.\footnote{More details on the numerical process can be found in \cite{Basu:2011yg}.} The results are consistent with a $\rho-\mu$ relationship of the form
\be
\rho=c_{sol}(q,m^2)\mu^4.
\ee
As in the probe case, the effective higher dimension of the space dictates the power in the relationship. The dependence of $c_{sol}(q,m^2)$ on $q$ and $m^2$ is as in the backreacted black hole case, section \ref{hairy}. Like the black hole with scalar field, the backreacted soliton with scalar field gives the same dominant power $\alpha$ as the corresponding probe case.

\section{Discussion}\label{conc}

In these notes we studied the $\rho-\mu$ relation for a variety of holographic field theories, and set conditions for physically consistent relationships based on local stability and causality. We observed that all of the examples considered are well modelled by a power law $\r=c\m^\a$ in the large $\m$ regime and that none of them fail to satisfy any of the general constraints stablished in sections \ref{gaugefield} and \ref{bg}. Except for the case of a probe gauge field in the Schwarzschild-AdS black hole background, the power $\alpha$ in all the bottom-up models obeyed the generic dimensional argument discussed in the introduction, as can be seen in table \ref{summary2}. This resulted in a larger power for the models with an extra periodic dimension. The brane constructions, table \ref{summary1}, displayed a larger variety of power laws, with the range $1<\a\le 3$, where $\alpha$ depended on the particular dimensions of the probe and background branes.

The study of bottom-up models led to the conclusion that, in general, the presence of a non-zero profile for the scalar field in the bulk induces a larger charge density on the boundary. In most cases, this change was realized as an increase of the scaling coefficient $c$ while the power law was kept unaltered. The only exception was the probe Einstein-Maxwell case, section \ref{phen_probe}. Here, in the absence of a scalar field, the probe Maxwell field enjoys its standard linear equations of motion, and naturally we find a linear $\r-\m$ relationship. With a non-zero scalar field, the power law becomes $\rho\propto\m^{d-1}$, as expected for the underlying CFT. In systems with an extra periodic direction, the numerical results displayed in figure \ref{solbhcompa} support the conclusion that at a given (large enough) chemical potential, the solitonic phase is denser than the corresponding black hole phase.

Despite our attempt to survey a large variety of holographic models, we do not claim to have presented a complete report and we do not discard the possibility of finding different $\r-\m$ relations in other types of bottom-up and top-down models. For example, one generalization would be to include $N_f>1$ flavour branes in the D$p$-D$q$ systems; this has been shown to change the power $\alpha$ in the relation \cite{Rozali:2007rx}. It would be interesting to extend this study to other classes of systems and to see how the results compare to those given here.

\section*{Acknowledgements}\label{acknow}

We thank Michael McDermott and Kevin Whyte for useful comments on the manuscript and Josh Davis, Joanna Karczmarek, and especially Mark Van Raamsdonk for helpful discussions. This work is supported in part by the Natural Sciences and Engineering Research Council of Canada.

\bibliographystyle{naturemag}
\bibliography{chemicalpotentialrefs}

\end{document}